\documentclass[twoside,twocolumn,9pt]{article}
\usepackage{extsizes}
\usepackage[super,sort&compress,comma]{natbib} 
\usepackage[version=3]{mhchem}
\usepackage[left=1.5cm, right=1.5cm, top=1.785cm, bottom=2.0cm]{geometry}
\usepackage{balance}
\usepackage{times,mathptmx}
\usepackage{sectsty}
\usepackage{graphicx} 
\usepackage{lastpage}
\usepackage[format=plain,justification=justified,singlelinecheck=false,font={stretch=1.125,small,sf},labelfont=bf,labelsep=space]{caption}
\usepackage{float}
\usepackage{fancyhdr}
\usepackage{fnpos}
\usepackage[english]{babel}
\addto{\captionsenglish}{%
  
}
\usepackage{array}
\usepackage{droidsans}
\usepackage{charter}
\usepackage[T1]{fontenc}
\usepackage[usenames,dvipsnames]{xcolor}
\usepackage{setspace}
\usepackage[compact]{titlesec}
\usepackage{hyperref}

\usepackage{epstopdf}

\definecolor{cream}{RGB}{222,217,201}

\usepackage[final]{changes}
\setdeletedmarkup{}
\definechangesauthor[name=Mathieu, color=red]{MG}
\definechangesauthor[name=Sergio, color=red]{SDT}

\begin{document}

\pagestyle{fancy}
\thispagestyle{plain}
\fancypagestyle{plain}{

\renewcommand{\headrulewidth}{0pt}
}

\makeFNbottom
\makeatletter
\renewcommand\LARGE{\@setfontsize\LARGE{15pt}{17}}
\renewcommand\Large{\@setfontsize\Large{12pt}{14}}
\renewcommand\large{\@setfontsize\large{10pt}{12}}
\renewcommand\footnotesize{\@setfontsize\footnotesize{7pt}{10}}
\makeatother

\renewcommand{\thefootnote}{\fnsymbol{footnote}}
\renewcommand\footnoterule{\vspace*{1pt}%
\color{cream}\hrule width 3.5in height 0.4pt \color{black}\vspace*{5pt}} 
\setcounter{secnumdepth}{5}

\makeatletter 
\renewcommand\@biblabel[1]{#1}            
\renewcommand\@makefntext[1]%
{\noindent\makebox[0pt][r]{\@thefnmark\,}#1}
\makeatother 
\renewcommand{\figurename}{\small{Fig.}~}
\sectionfont{\sffamily\Large}
\subsectionfont{\normalsize}
\subsubsectionfont{\bf}
\setstretch{1.125} 
\setlength{\skip\footins}{0.8cm}
\setlength{\footnotesep}{0.25cm}
\setlength{\jot}{10pt}
\titlespacing*{\section}{0pt}{4pt}{4pt}
\titlespacing*{\subsection}{0pt}{15pt}{1pt}

\renewcommand{\headrulewidth}{0pt} 
\renewcommand{\footrulewidth}{0pt}
\setlength{\arrayrulewidth}{1pt}
\setlength{\columnsep}{6.5mm}
\setlength\bibsep{1pt}

\makeatletter 
\newlength{\figrulesep} 
\setlength{\figrulesep}{0.5\textfloatsep} 

\newcommand{\topfigrule}{\vspace*{-1pt}%
\noindent{\color{cream}\rule[-\figrulesep]{\columnwidth}{1.5pt}} }

\newcommand{\botfigrule}{\vspace*{-2pt}%
\noindent{\color{cream}\rule[\figrulesep]{\columnwidth}{1.5pt}} }

\newcommand{\dblfigrule}{\vspace*{-1pt}%
\noindent{\color{cream}\rule[-\figrulesep]{\textwidth}{1.5pt}} }

\makeatother


\title{\textbf{Fission of charged nano-hydrated ammonia clusters - microscopic insights on the nucleation processes}} 
\author{Bart Oostenrijk,$^{\ast}$\textit{$^{a}$} Dar\'io Barreiro,\textit{$^{b\ddag}$} Noelle Walsh,\textit{$^{a}$} Anna Sankari,\textit{$^{a}$} Erik P. M\aa nsson,\textit{$^{c}$} Sylvain Maclot,\textit{$^{a,d}$} Stacey Sorensen,\textit{$^{a}$}
\\ Sergio D\'iaz-Tendero,\textit{$^{b,e,f}$} and Mathieu Gisselbrecht\textit{$^{a}$}}
\date{}

\maketitle 

\abstract{While largely studied on the macroscopic scale, the dynamics leading to nucleation and fission processes in atmospheric aerosols are still poorly understood at the molecular level.
Here, we present a joint experimental-theoretical study of a model system consisting of hydrogen-bonded ammonia and water molecules. Experimentally, the clusters were produced via adiabatic co-expansion. Double ionization ionic products were prepared using synchrotron radiation and analyzed with coincident mass- and 3D momentum spectroscopy.  Calculations were carried out using \textit{ab initio} molecular dynamics to understand the fragmentation within the first $\sim$500 fs.  Further exploration of the potential energy surfaces were performed at a DFT level of theory to gain information on the energetics of the processes.
Water was identified as an efficient nano-droplet stabilizer, and is found to have a significant effect even at low water content. On the molecular level, the stabilizing role of water can be related to an increase in the dissociation energy between ammonia molecules and the water enriched environment at the cluster surface. Furthermore, our results support the role of ammonium as a charge carrier in the solution, preferentially bound to surrounding ammonia molecules, which can influence atmospheric nucleation process.
} \\


\renewcommand*\rmdefault{bch}\normalfont\upshape
\rmfamily
\section*{}
\vspace{-1cm}

\footnotetext{\textit{$^{a}$~Department of Physics, Lund University, Box 118, 22100 Lund, Sweden.; E-mail: mathieu.gisselbrecht@sljus.lu.se}}
\footnotetext{\textit{$^{b}$~Departamento de Qu\'imica, M\'odulo 13, Universidad Aut\'onoma de Madrid, 28049 Madrid, Spain }}
\footnotetext{\textit{$^{c}$~Attosecond Science group, DESY Photon Science Division, Schenefeld, Germany. }}
\footnotetext{\textit{$^{d}$~Biomedical and X-Ray Physics, Department of Applied Physics, AlbaNova University Center, KTH Royal Institute of Technology, SE-10691 Stockholm, Sweden}}
\footnotetext{\textit{$^{e}$~IFIMAC, Universidad Aut\'onoma de Madrid, 28049, Madrid, Spain}}
\footnotetext{\textit{$^{f}$~IAdChem, Universidad Aut\'onoma de Madrid, 28049, Madrid, Spain}}





\section{Introduction}
Nucleation processes play a key role in the formation of aerosols in the atmosphere. Though these processes  can be largely explained on the macroscopic scale, the very details of mechanisms occurring on the molecular level have also been shown to be relevant \cite{Kulmala_2013}. In particular, in the troposphere, molecular ionization (e.g. by cosmic background radiation \cite{Yu_2001}) is proposed to have a large influence on the formation of aerosol particles \cite{Lee2003,Curtius2007}, and is possibly important in the formation of clouds \cite{Arnold_1982, Arnold2008, Kazil2008} involving (mainly) water molecules \cite{Salby_1996}. A crucial step towards a better understanding of the role of molecular ionization in nucleation processes requires bridging the gap between isolated molecules and stable clusters ($<$ 2 nm) in the description of these processes. Due to the increasing abundance of ammonia in the earth's troposphere \cite{Ziereis_1985, Hopfner_2016} (originating from anthropogenic sources \cite{Bouwman_1997}), we focus our investigation on multiply-charged nano-hydrated ammonia clusters.
Such particles might act as a nucleation enhancer, thus explaining the high particle formation rates of atmospheric aerosols containing sulfuric acid \cite{Ball1999,Torpo2007,Almeida2013,Jonas_2017}.

\begin{figure}[b!]
\centerline{\includegraphics[width=0.5\textwidth]{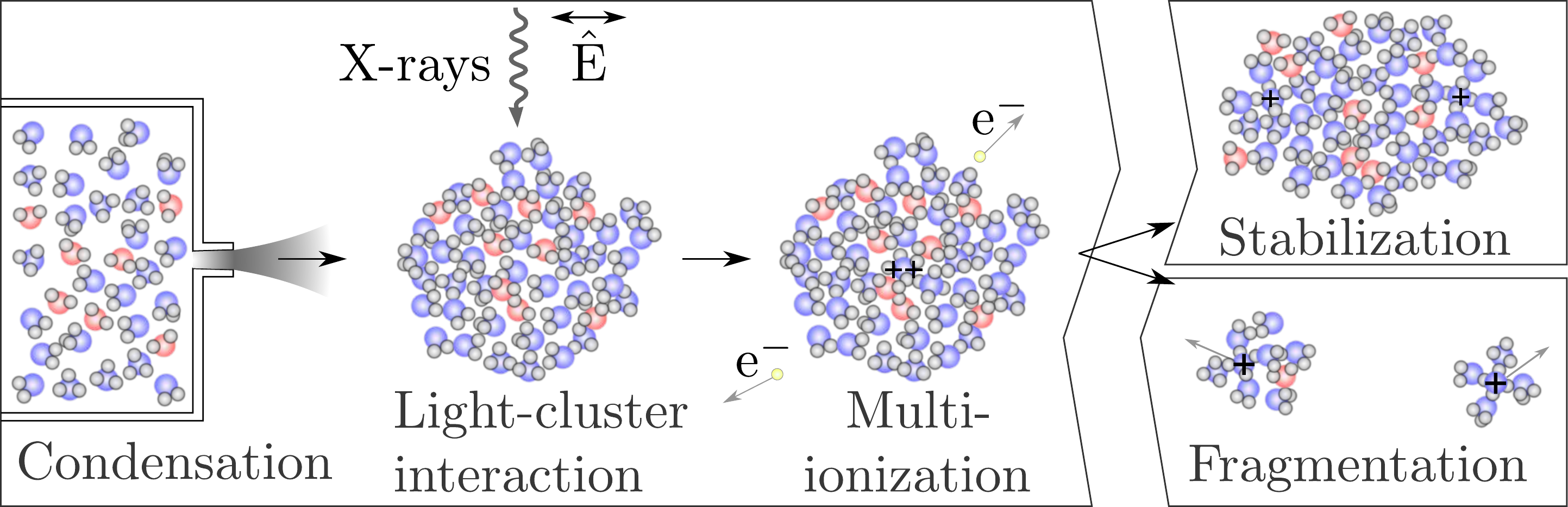}}
\caption{Schematic representation of the experiment. From left to right: co-expanded ammonia -and water- molecules are condensed to neutral clusters, they interact with X-rays (with polarization \^{E}). The interaction can lead to double ionization of the cluster. In this case, the irradiated cluster will either accomodate both charges, forming a stable dication, or dissociate into smaller charged fragments.}
\label{fig:concept}
\end{figure}

\begin{figure*}
\centering
 \includegraphics[width=0.9\textwidth]{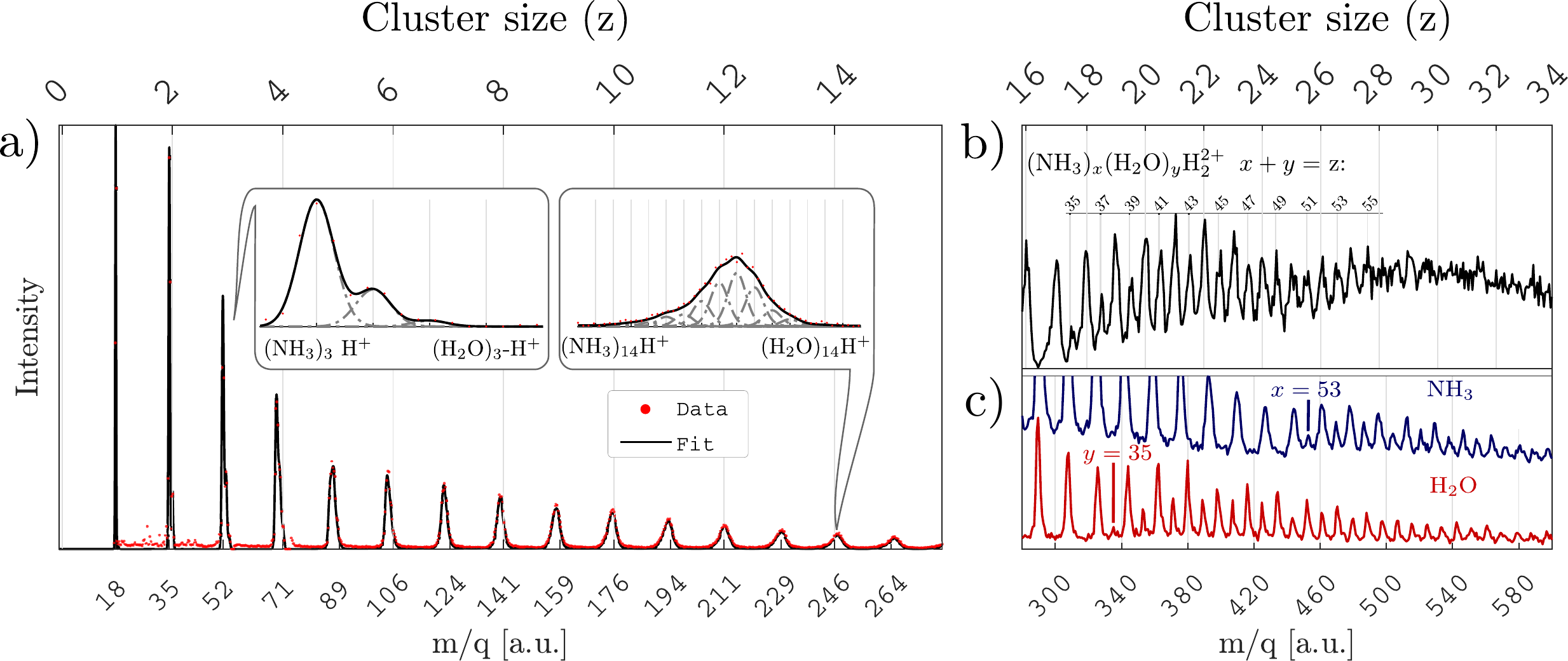}
 \caption{\added[id=MG]{Mass-to-charge spectra upon ionization of mixed water-ammonia clusters (a,b) and pure clusters (c). $x$ denotes the number of $\ce{NH_3}$  molecules in the cluster, $y$ denotes the number of $\ce{H_2O}$  molecules in the cluster and for mixed clusters, the total number of molecules, $z=x+y$. (a) The projection of the double-coincidence mass spectrum recorded (red dots) at $h \nu =$ 450 eV, 400 mbar stagnation nozzle pressure, and a water pressure fraction $\frac{P_{H_2O}}{P}$ of 13 \%. The composition of protonated fragments have been determined by fitting individual peaks for singly charged ions, $\ce{(NH_3)_x (H_2O)_y H}^{+}$  . For the resulting fit to each peak (black lines) see insets. (b) The single-coincidence mass spectrum where the cluster composition is described by $\ce{(NH_3)_x (H_2O)_y H}^{2+}_{2}$. The onset (appearance size) of stable mixed doubly charged clusters is observed at z=35. c) For comparison, our experimentally determined dication appearance sizes for pure ammonia (blue), x=53, and pure water (red), y=35, are indicated in the colour-coded mass spectra. The pure ($\ce{NH_3}, \ce{H_2O}$) cluster spectra originate from $h \nu=$ 420 and 550 eV at $P=$ 200 and 680 mbar, respectively.}}
\label{fig:m2q}
\end{figure*}

For these studies, we implemented an advanced multi-coincidence 3D momentum imaging mass spectrometry technique to unravel fundamental mechanisms behind ionization-induced charge and proton dynamics. These dynamics can occur at different time and length scales in hydrogen-bonded clusters of 1-2 nm size\cite{Oostenrijk_2018}. The principle of the experiment is illustrated in Fig. \ref{fig:concept}. Adiabatic cooling of water and ammonia vapors is achieved via co-expansion in vacuum causing molecules to condense into nano-hydrated ammonia clusters, \ce{(NH_3)_m (H_2O)_n}. These clusters absorb a single X-ray photon and in most cases this leads to the formation of doubly-charged species. Depending on the original neutral cluster size, a dicationic cluster can either remain stable or dissociate with charge separation. Coincidence techniques enable the measurement of all ionic fragments originating from a single parent cluster, allowing us to follow the water and ammonia content in the cluster as a function of the fragment size and ultimately identify the molecular sites in the cluster that act as 'charge host'. The 3D momentum imaging technique enables the characterization of the kinematics of fragmentation and even to trace interactions between molecules.  

In this work, we report on the first study of the stability and fragmentation dynamics of doubly-charged mixed water-ammonia clusters. We employ synchrotron radiation with photon energies ranging from 60 eV to 550 eV to explore effects related to ionization and the subsequently induced fragmentation dynamics. We observe that the stability of protonated doubly-charged clusters depends on the ammonia concentration. The formation of a shell-like structure around the charge host in the cluster \cite{Payzant_1973, Nakai2000201, Meot-ner_1986} is confirmed. The elementary constitution of this shell-like  structure can be changed by both the initial concentration of water in the cluster and the excess energy deposited by high energetic photons. Finally, our data reveals that the interaction of ionic ammonia molecules with surrounding water molecules influences the kinematics of fragmentation. Further insights on the nucleation and fragmentation after charge separation and migration are obtained with quantum chemistry calculations.
\section{Methods}
\subsection{Experiments}
The measurements were carried out at the soft X-ray beamline I411 at MAX-lab Lund, Sweden \cite{BASSLER1999}. The multi-ion coincidence 3D momentum imaging spectrometer used in this work was described in detail elsewhere \cite{Laksman2013}. In short, an electron emitted after photo-absorption is detected and initiates the recording of an event. The mass and linear momentum of ions that arrive at the detector within a 30 $\mu$s acquisition time window are registered. Care was taken to measure within a clean `coincidence regime'.\added[id=MG]{, i.e. the probability of measuring a false coincidence is about 3-6\%. In this regime, the recorded particles in one event essentially originate from the dissociation/fragmentation of the same `parent' cluster, there no need to subtract random events. The treatment of coincidence data for cluster experiments are discussed in a recent publication\cite{Oostenrijk_2018}. In the following we will refer to single-coincidence data as events where only one ion was detected within our acquisition time window of 30 $\mu$s. Likewise  double-coincidence data refers to events where two ions were detected within the acquisition time window. Note that the detection efficiency (around 35\%) decreases at higher masses \cite{Gilmore_2000}, and causes an underestimation of the abundance of larger clusters.} For the measurement, the spectrometer axis was oriented perpendicular to the polarization vector of the light. 

The clusters were produced in a supersonically expanded jet through a conical nozzle (20$^{\circ}$ full opening angle, 150 $\mu$m throat), and doubly skimmed before entering the spectrometer. A steady flow of ammonia (Linde gas, $<0.001$ permille impurity) was directed through a liquid water container, without using a buffer gas. The container and expansion nozzle were kept at 34$^{\circ}C$, corresponding to a water partial pressure of 53 mbar \cite{CRC_handbook_2004}. The pressure of the gaseous ammonia was varied between 290 and 720 mbar. The relative contributions of water and ammonia molecules in cluster formation at larger sizes (above 8 units) are unknown from literature, but the pure ammonia neutral cluster size distribution model estimates an average size of approximately 30 and 110 ammonia molecules at 237 and 667 mbar, respectively \cite{Bobbert_2002}.
We denote the different stagnation conditions by the total stagnation pressure ($P$) and the relative water pressure $\left( \frac{P_{H_2O}}{P} \right)$.


\added[id=MG]{Fig. \ref{fig:m2q} a) presents the double-coincidence mass spectrum of cluster products arising from the break up with charge separation of unstable dicationic ammonia-water clusters into two smaller charged fragments after soft X-ray ionization. At this photon energy, almost all cluster products are expected to have undergone proton transfer and are observed as protonated species just like in the cases of pure ammonia and pure water clusters. Hence it is possible to unambiguously assign the mass of a product to its composition $\ce{(NH_3)_x (H_2O)_y H}^{+}$, where $x$ and $y$ denotes the number of $\ce{NH_3}$ and $\ce{H_2O}$ molecules, respectively. We can then define the total number of molecules, $z=x+y$, and , at a given z value, the masses of ionic fragments with different compositions  are only separated by one atomic mass unit (17 and 18 a.u. for \ce{NH_3} and \ce{H_2O}, respectively). Since metastable evaporation occurs on a microsecond timescale \cite{Wei_1990, Belau_2007}, about four times slower than the acceleration time in the spectrometer, we can also assume that the mass of the cluster product in the mass spectra reflects its mass in the interaction region before any metastable evaporation\cite{Oostenrijk_2018}. Hence, the total number of molecules, $z$, can be interpreted as the minimum cluster size of the dicationic clusters just after break up.\\
Since the peak width induced by the ion kinetic energy causes an overlap of the neighbouring cluster mass peaks we performed a fit of multiple Gaussian peaks to recover the individual composition intensities. The resulting fit is presented as a black line in Fig. \ref{fig:m2q}a, with detailed views in the insets. By combining the composition of each individual peak and its intensity, the relative concentration of water for a given cluster size, $z$, can be determined. Note that the small percentage of cluster products which remains unprotonated has a negligible impact on the analysis.\\
At higher cluster sizes, the doubly charged clusters are stable, i.e. they do not undergo fragmentation with charge separation, and the appearance of these dications can be seen in Fig. \ref{fig:m2q} b). The dications are not observed in double- or multiple-ion coincidence spectra, which imply that no breakup with charged species occurred while fast evaporation of molecules (timescale $<$ a few nanosecond) is possible after proton transfer. The first stable doubly charged (doubly protonated) clusters is observed at $z = 35$ ($z=x+y$, \ce{(NH_3)_x (H_2O)_y H}$_2^{2+}$), and compared to the dication appearance sizes of both pure ammonia and pure water clusters, at 53 and 35 units, respectively in Fig. \ref{fig:m2q} c).}

\subsection{Computational Details}
In order to give insight into the experimental measurements, we have studied the behavior of mixed water-ammonia clusters with simulations based on quantum chemistry methods. To this end, we have considered different cluster sizes: (NH$_3$)$_x$(H$_2$O)$_y$ with ($x,y$) = (2,8), (5,5), (8,2), (6,18), (11,12), (18,6), (10,40), (25,25) and (40,10). To find the most stable neutral conformers for these clusters, semi-classical calculations using Self-Consistent Charge Density-Functional Tight Binding based calculations (DFTB-SCC) were carried out\cite{Elstner1998} applying the Parallel Tempering Molecular Dynamics (PTMD) formalism \cite{SUGITA1999,Oliveira2015}. These calculations were performed using the deMon-Nano program\cite{demonnano}, in particular a modified version that includes specific water-ammonia interaction potentials developed by Cuny et al.\cite{Simon2018}. DFTB is a tight-binding method parametrized using Density Functional Theory (DFT); it is not an \textit{ab initio} method since it contains certain empirical parameters, although most of them introduced within solid theoretical basis. Despite being less accurate than DFT, the computational scaling is much less limited for big systems, therefore it allows to reach longer time scales in molecular dynamics \cite{KOSKINEN2009,Seifert2007}. Furthermore, the SCC extension helps to avoid the uncertainties obtained within the non-SCC approach in systems where the charge constitutes a main factor in the chemical bonding \cite{Elstner1998,Frauenheim2000,KOSKINEN2009,Seifert2007}. On the other hand, PTMD is a powerful tool to explore the potential energy surface (PES) of systems with a large number of degrees of freedom\cite{Nymeyer2008}. Starting from different temperatures, several trajectories run in parallel allowing exchange of the structures at given points of the trajectories between the considered temperatures \cite{Earl2005}, thus exploring separate regions on the PES and locating non-connected minima. We performed DFTB-SCC simulations with the PTMD formalism in the above mentioned clusters including 32 parallel trajectories with temperatures in the range 200 - 1000 K and propagation up to t = 300 ps with a time step of $\Delta$t = 0.3 fs; every 30 fs random structure exchange over the parallel trajectories is performed. From these trajectories we selected the 10 structures with lowest energy in the PES, and we optimized their geometry at the DFTB-SCC level. In all the DFTB-SCC calculations, atomic populations were computed with the CM3 model \cite{Li1998}, and dispersion and Coulombic interactions were corrected by the method proposed by Rapacioli et al. \cite{Rapacioli2009}.

\begin{figure}[t!]
\centering
 \includegraphics[width=0.5\textwidth]{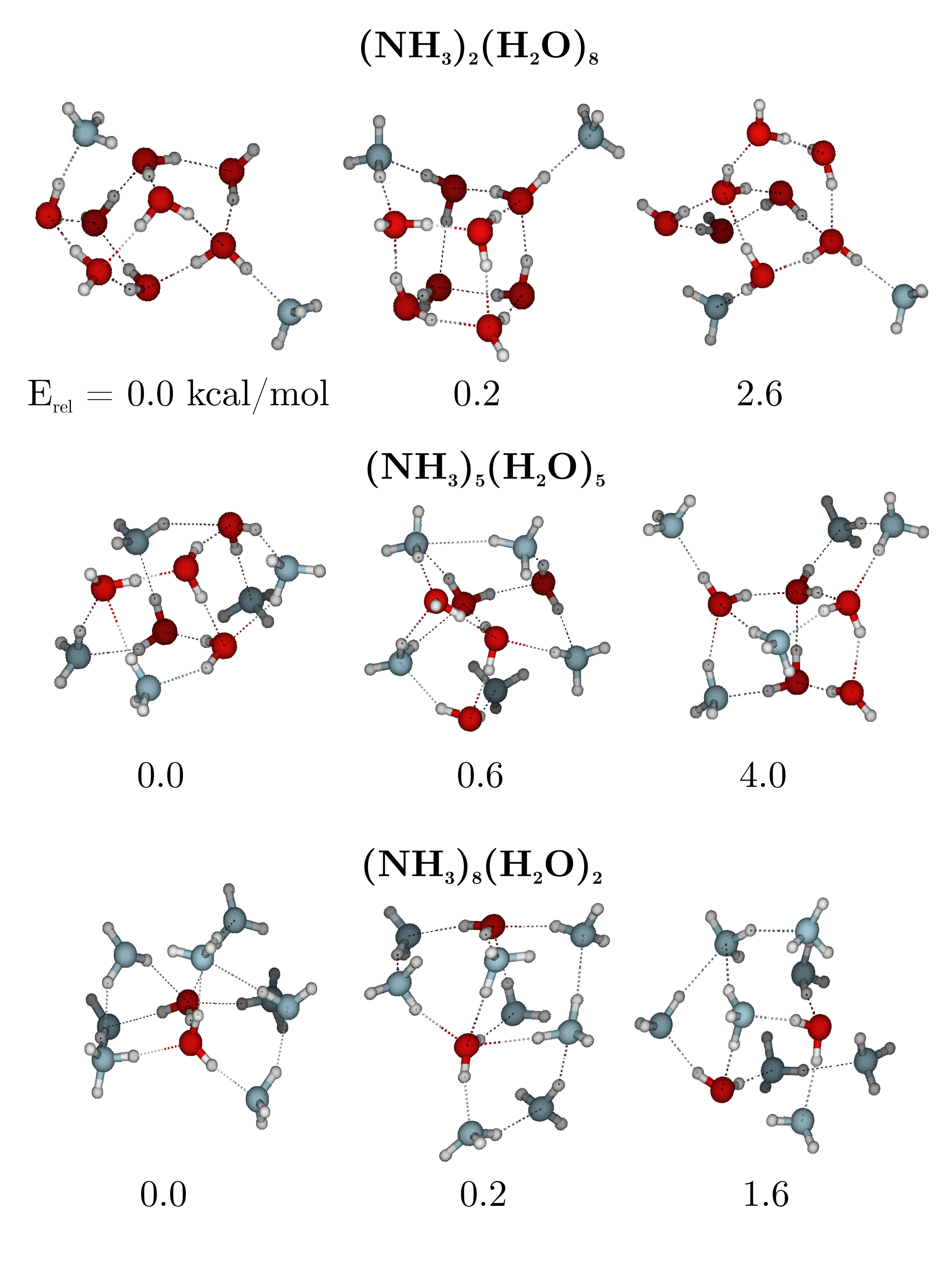}
 \caption{: Geometries of the 3 most stable conformers for each combination chosen for the 10 member nano-hydrated ammonia clusters. The number below each structure indicates the relative energy between each set of clusters in kcal/mol.}
\label{fig:str10}
\end{figure}

\begin{figure}[t]
\centering
 \includegraphics[width=0.5\textwidth]{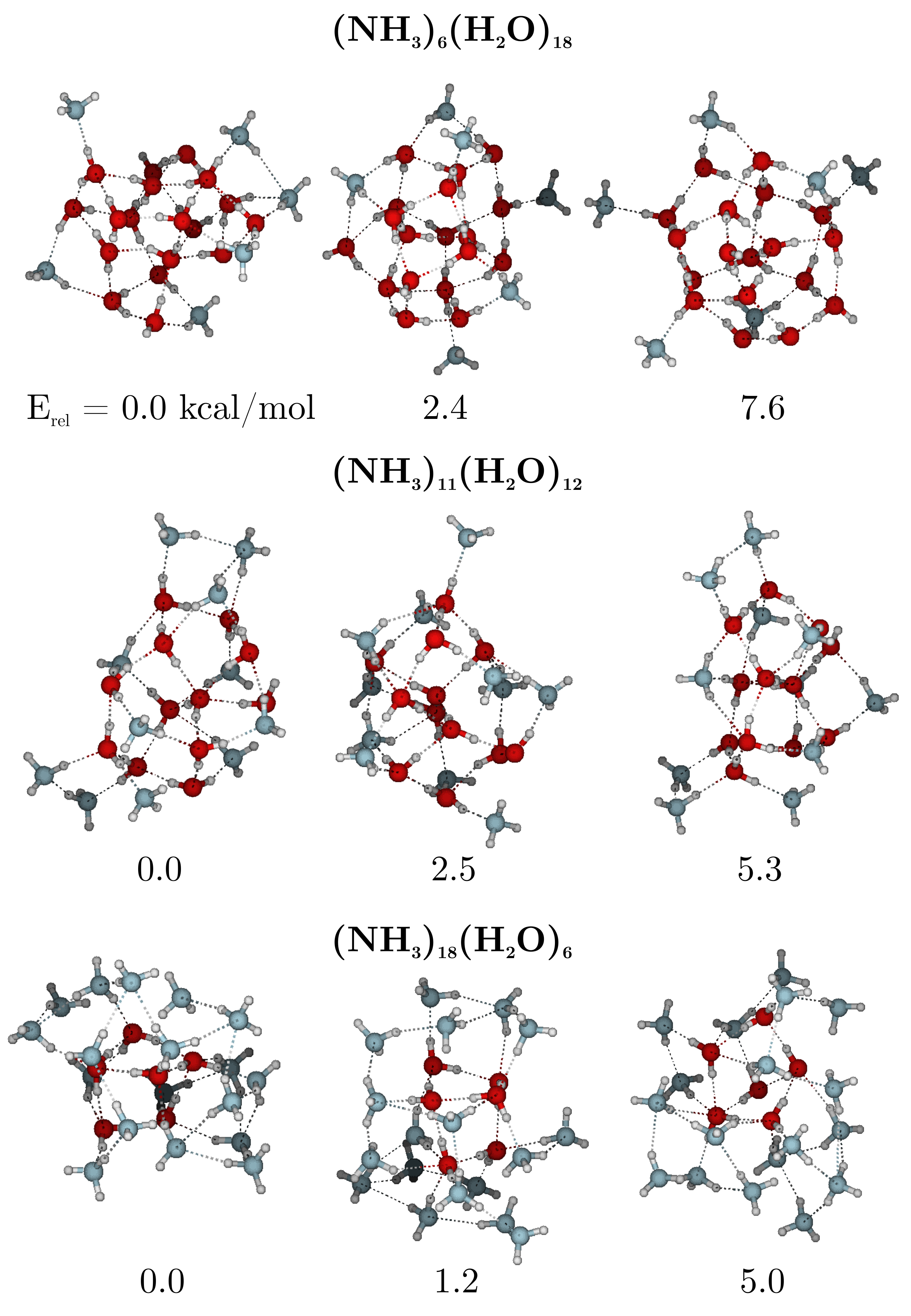}
 \caption{: Geometries of the 3 most stable conformers for each combination chosen for the 24 member nano-hydrated ammonia clusters. The number below each structure indicates the relative energy between each set of clusters in kcal/mol.}
\label{fig:str24}
\end{figure}

\begin{figure*}
\centering
 \includegraphics[width=1\textwidth]{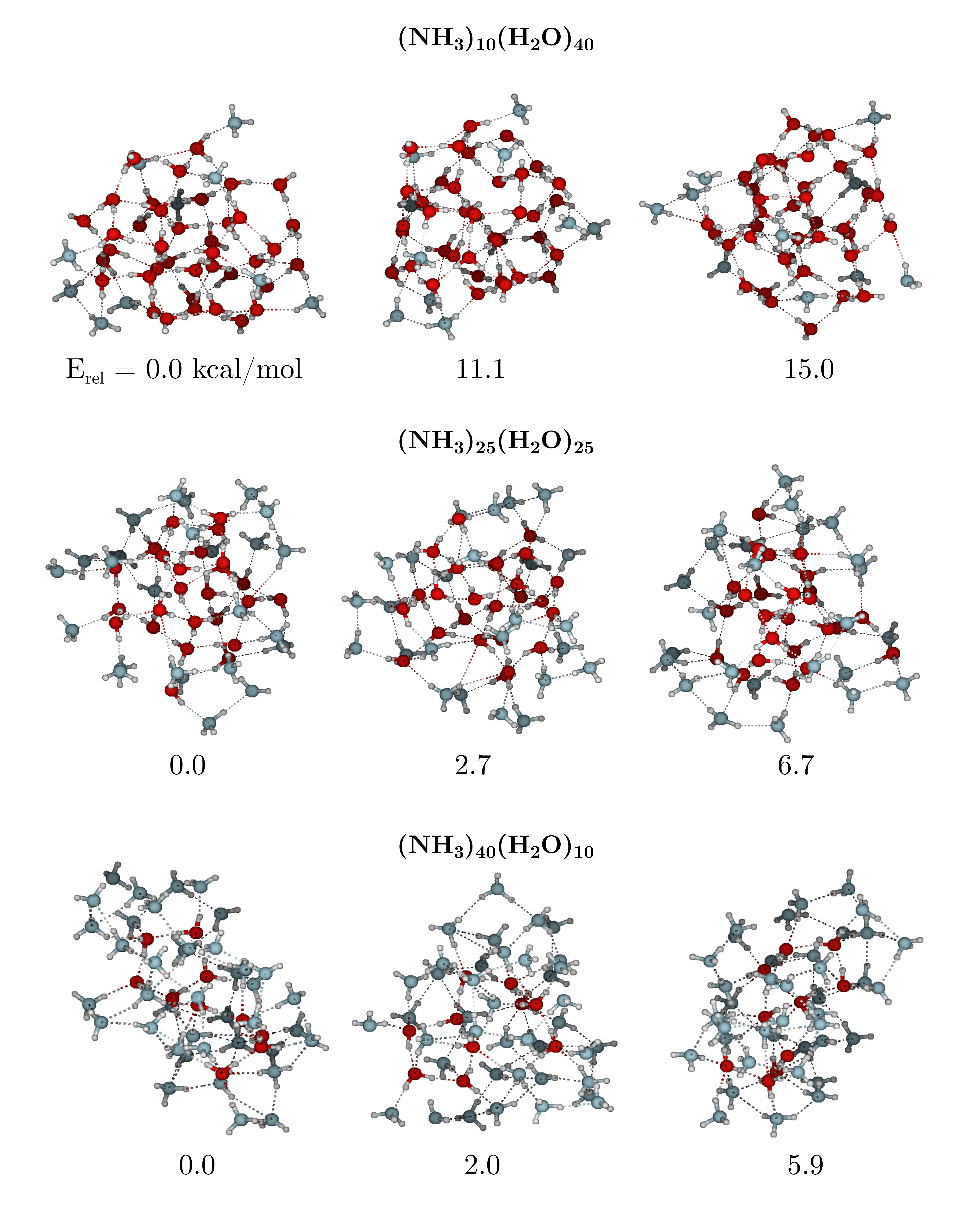}
 \caption{: Geometries of the 3 most stable conformers for each combination chosen for the 50 member nano-hydrated ammonia clusters. The number below each structure indicates the relative energy between each set of clusters in kcal/mol.}
\label{fig:str50}
\end{figure*}

All the DFTB-SCC optimized structures were afterwards reoptimized at a DFT level of theory using the M06-2X functional, which is a meta-hybrid GGA functional that includes part of Hartree-Fock exchange and it was parametrized only for non-metals\cite{Zhao2008}. This functional was used in combination with the 6-31++G(d,p) basis set in the geometry optimization. A selection of the most stable structures obtained is shown in Fig. \ref{fig:str10}, \ref{fig:str24} and \ref{fig:str50}.

\indent The next step of our computational strategy consists in performing molecular dynamics simulations on ionized and excited clusters using the optimized geometry of the neutral clusters as initial structure. This approach relies on a sudden ionization of the neutral clusters in a Franck-Condon way, in particular we considered doubly positively charged clusters.  To mimic the experimental conditions, we have introduced a certain amount of excitation internal energy E$_{\rm exc}$, randomly redistributed among the nuclear degrees of freedom. The \textit{ab initio} molecular dynamics calculations were performed using the atom-centered Density Matrix Propagation (ADMP) method. Trajectories have been carried out at a DFT level of theory, also using the M06-2X functional with the 6-31++G(d,p) basis set. All simulations were performed with a time step of $\Delta$t = 0.1 fs, a fictitious mass of $\mu$ = 0.1 a.u. and a maximum simulation time of t$_{\rm max}$ = 300 fs. In such large molecular clusters, the dynamics of excited electronic states can be simulated assuming that statistical processes redistribute internal electronic energies as heat over all nuclear degrees of freedom. A typical excess/excitation internal energy, E$_{\rm exc}$, ranging from 10 to 20 eV, was used to consider states prepared upon double ionization, e.g. after Auger decay. For each conformer and each value of excitation energy 100 trajectories have been carried out. The simulations were performed starting from the most stable conformer of the 10 unit clusters: $\rm (NH_3)_8(H_2O)_2$, $\rm (NH_3)_5(H_2O)_5$ and $\rm (NH_3)_2(H_2O)_8$; since we can consider that the nature of the reaction pathways of all cluster sizes must be identical to that of larger systems. Statistics were performed over these trajectories. Finally, we identified the dominant processes observed in the dynamics and from them, we obtained final dissociation energies of some selected fragmentation channels. They were computed at the same level of theory: DFT - M06-2X/6-31++G(d,p). All DFT calculations were performed using the Gaussian09 program \cite{Gaussian09}.

\section{Results \& Discussion}
\begin{figure}[t]
\centerline{\includegraphics[width=0.5\textwidth]{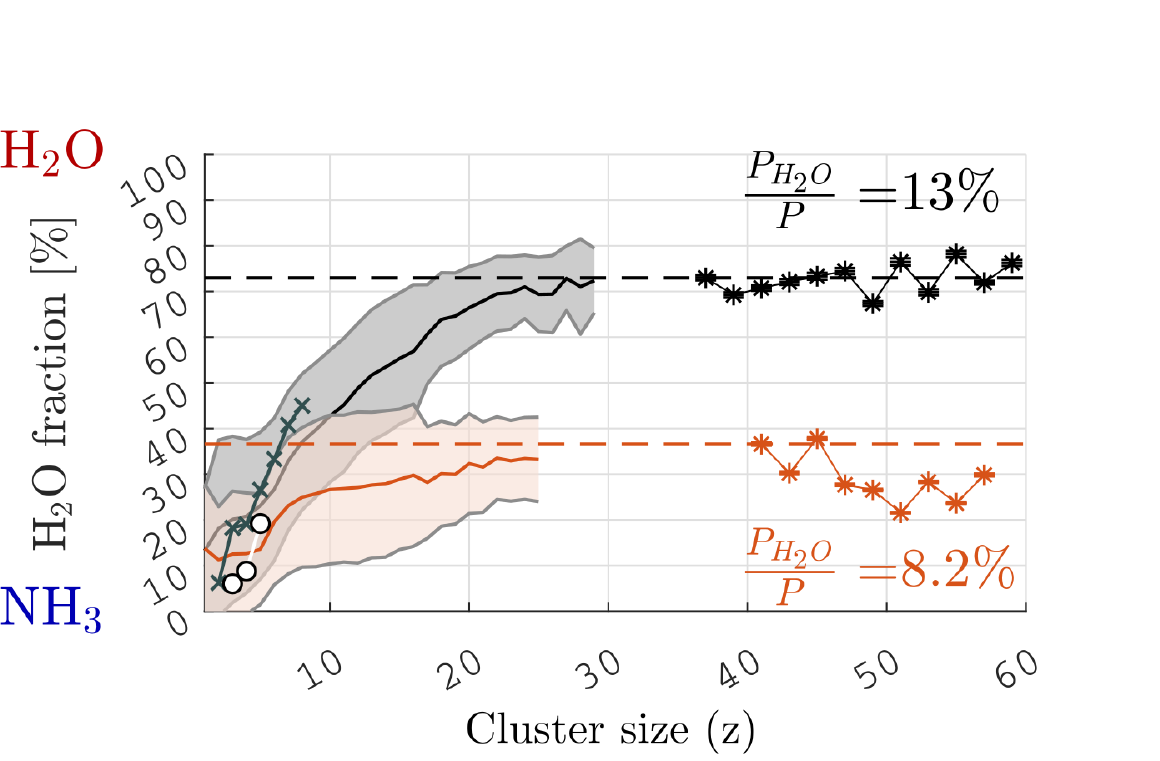}
}{\caption{The average cluster water fraction (solid lines), and standard deviation (areas) at $h\nu=$ 450 eV and $\frac{P_{H_2O}}{P}=$ 13\% (black) and 8\% (orange), from fits as shown in Figure \ref{fig:m2q} a), along with the fraction of doubly charged clusters (asterisks) and the water fraction of the smallest dications as dashed lines. Water fractions from multi-photon ionization \cite{Choo_1983} and alpha irradiation \cite{Hogg_1965} are shown in crosses and circles, at $\frac{P_{H_2O}}{P}=$ 9.1\% and 75\%, respectively.}
\label{fig:comp}
}
\end{figure}

The transition between macroscopic and microscopic behaviour can be investigated following the ionization of mixed water-ammonia nanodroplets by soft X-ray radiation. With photon energies exceeding multiple ionization thresholds, clusters with two or more charges are produced. Multi-coincidence 3D-momentum imaging technique allows us to selectively focus on doubly charged clusters - both stable doubly charged ions and dicationic clusters breaking apart into small singly-charged fragments.
 Fig. \ref{fig:comp} shows the average water-ammonia composition of both the singly charged fragments (solid lines) and the stable dication clusters (asterisks), at two different stagnation mixtures. For clusters ranging in size from a few molecules to those with diameters of about 1-2 nm, we find that the water content follows a monotonically increasing function, and stabilizes after a few tens of molecules to an asymptotic value close to that of the stable dications. Generally, in a macroscopic binary solution, the relative concentration of each compound is independent of the size of the sample. Our result shows that this statement does not hold at the microscopic scale, due to molecular interactions that lead to local composition changes. The water mixing in the smallest fragments is limited to a relative fraction of about 30\% or lower, indicating that most fragments contain at least one ammonia, most likely in protonated form (ammonium, \ce{NH_4^+}, \cite{Hogg_1965}) as a stable host of a single charge.\\
For the comparison of the size dependent composition observed in our measurements, Fig. \ref{fig:comp} includes further measurements reported in literature that were performed at 9.1\% \ce{H_2O} fraction using `soft' two-photon ionization (total 12.8 eV) \cite{Choo_1983}, shown as crosses, and at a 75\% \ce{H_2O} relative water pressure using alpha irradiation \cite{Hogg_1965}, shown as white circles. In all cases, a similar composition of the small fragments is observed, independent of the ionization mechanisms and expansion conditions, indicating an efficient production of small fragments \added[id=MG]{(z$\leq$ 5)} with a high ammonia concentration.  The clear propensity to form a high ammonia concentration in small fragments up to five molecular units \added[id=MG]{($z=5$)} supports the anticipated abundance of ammonia ligands around the charged ammonium core \cite{Choo_1983, Hogg_1965, Payzant_1973}. The onset of water mixing around the size of $z=$3-5 molecular units has recently been explained as an ammonium (\ce{NH_4^+}) core taking up water ligands as well \cite{Hvelplund_2010}. \added[id=MG]{The remarkable agreement between all experimental studies suggests that mass spectrometry measures the end product of a complex fragmentation dynamics where reorganization and cooling down of the ionic clusters will be independent of the source to produce clusters, the initial charge states of the cluster and the breakup process}. 

\indent We will now investigate the products of the doubly charged nano-hydrated ammonia clusters starting from the general aspects of the more complex cluster and then moving into the processes observed into more simple environments.

\begin{figure} [t]
\centerline{\includegraphics[width=0.5\textwidth]{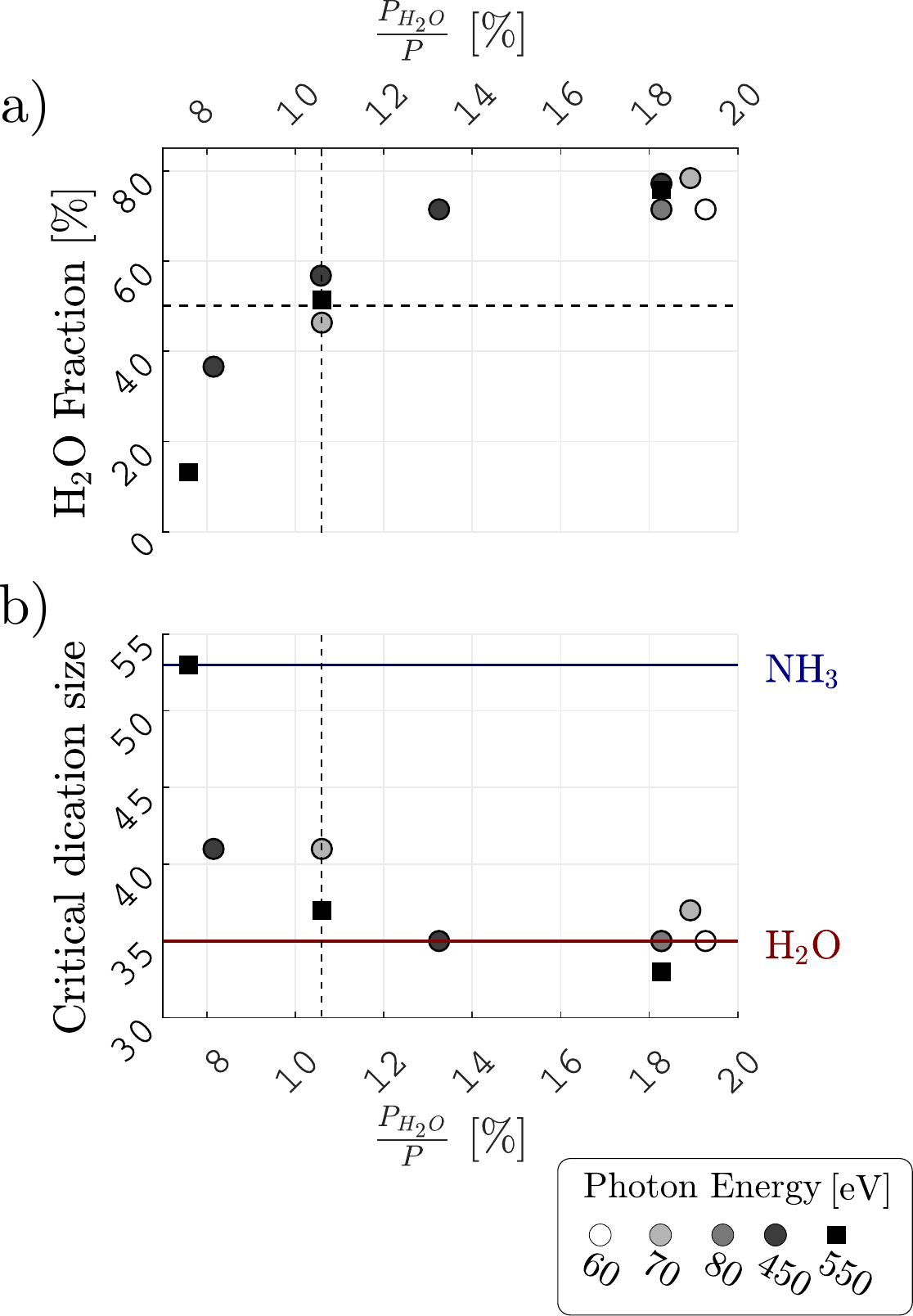}}
\caption{The critical dication water fraction (a), and critical dication appearance size (b), as a function of stagnation pressure ratios, at different photon energies indicated in the legend. Reference lines for pure ammonia (blue, $x=53$, reported at $x=51$ \cite{Shukla1984, COOLBAUGH198919}) and water (red, $y=35$, as reported \cite{Stace_1988}) dication appearance sizes.}
\label{fig:dication}
\end{figure}

\subsection{Nano-Liquid droplets}

Clusters that grow above a certain size can host two charges in a stable configuration. This size is defined as the critical dication size, or dication appearance size. The stability of clusters around the dication appearance size is often discussed in terms of the liquid drop model. This approach employs macroscopic properties to predict the stability, such as the surface tension and permittivity \cite{Rayleigh_1882,Casero_1988}, but the underlying reason for these properties can ultimately only be rationalized by studying the most stable structures and the interactions between molecules in an H-bonded lattice/network. 

\begin{figure}[h!]
\centerline{\includegraphics[width=0.5\textwidth]{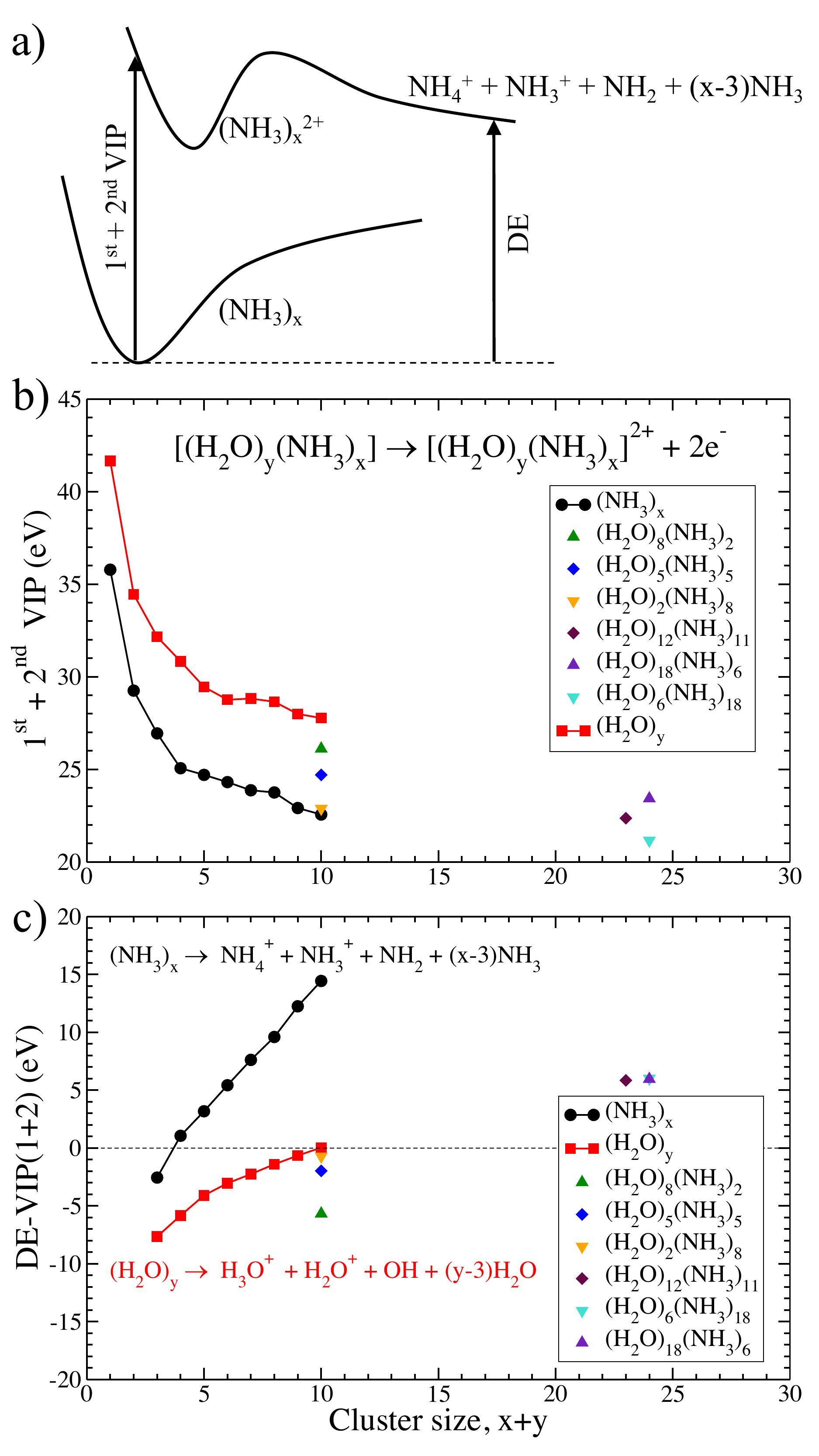}}
\caption{(a) Schematization of the ionization and dissociation processes studied for a (NH$_3$)$_x$ cluster. (b) First and second vertical ionization potentials (VIPs) calculated for pure and mixed water-ammonia clusters. (c) Lowest energy dissociation channels as a function of the 1$^{st}$ and 2$^{nd}$ VIPs for both pure and mixed nano-hydrated cluster. In the case of mixed clusters, the channel taken into consideration is:
${\rm (H_2O)_y(NH_3)_x \rightarrow y H_2O + NH_4^+ + NH_2 + NH_3^+ + (x-3)NH_3}$
except for ${\rm (H_2O)_8(NH_3)_2 \rightarrow 7 H_2O + NH_4^+ + NH_3OH^+}$.}
\label{fig:energies}
\end{figure}

In Fig. \ref{fig:dication}, we present the measured critical dication size  and composition at different stagnation conditions.
The critical dication size \added[id=MG]{(in Fig. \ref{fig:dication} b),} decreases non-linearly as the relative water pressure increases. The greater critical dication size of pure ammonia clusters can be understood by the lower surface tension of liquid ammonia with respect to liquid water ($\tfrac{\gamma_{\ce{NH_3}}}{\gamma_{\ce{H_2O}}} \approx 0.4$ \cite{CRC_Handbook_1989}). We note that the variation of the critical dication size is correlated to the composition in the cluster \added[id=MG]{(Fig. \ref{fig:dication} a)}. At an equal mix of water and ammonia in the cluster \added[id=MG]{(dashed-lines in Fig. \ref{fig:dication} a), }the critical size remains closer to the critical size of pure water clusters \added[id=MG]{(in Fig. \ref{fig:dication} b)}. Increasing the cluster water fraction to 80\% results in a critical dication size identical to that of pure water clusters. This trend can be understood using energetic arguments. First we have computed Vertical Ionization Potentials (VIPs) for pure water, pure ammonia and for the selected mixed clusters at a DFT level of theory (initial guess of the geometries before optimization for the pure ammonia cluster were taken from \cite{Malloum_2015} and for the pure water clusters from \cite{JAMES2005}). To perform these simulations we have taken the optimized geometry of the most stable neutral conformer and we have calculated the energy required to extract the two outer electrons in a Franck-Condon approximation keeping the geometry of the neutral species. At higher water fractions, clusters show higher 1$^{st}$+2$^{nd}$ VIPs as well as an exponential decay as the cluster size increases (Fig. \ref{fig:energies}b). Energetically, the mixed clusters lie in between the pure water and pure ammonia clusters. In a second step, we have computed dissociation energies including the zero point energy (ZPE) correction. In Fig. \ref{fig:energies}c we show dissociation channels where total dissociation has occurred, taking into account possible hydrogen transfers leading to the lowest dissociation energy channel with respect to the 1$^{st}$ and 2$^{nd}$ VIP. The most stable structure has been chosen in each case. Negative values mean that the dissociation energy lies below the first and second VIPs and therefore the channel would be favored. On the other hand, those values above zero, will require a redistribution of the excess internal energy in order to reach the dissociation channel. Pure water clusters start below this Coulomb limit while pure ammonia clusters lie above it. Observing those values for the mixed clusters it becomes evident that water fraction in the cluster has a direct effect on the variations in critical dication size. 
\added[id=SDT]{Even in the case of the $\rm (H_2O)_2(NH_3)_8$ cluster, with only two water molecules, their presence is enough to provide energies below the Coulomb limit. The reason is that the water molecules are located in the core of the cluster, enabling nucleation and thus keeping ammonia molecules strongly bonded.} Consequently  we can infer that the mixing of water effectively stabilizes the doubly charged cluster, more than expected from a linear behavior based on the water fraction in the cluster.

\begin{figure} [t]
\centerline{\includegraphics[width=0.5\textwidth]{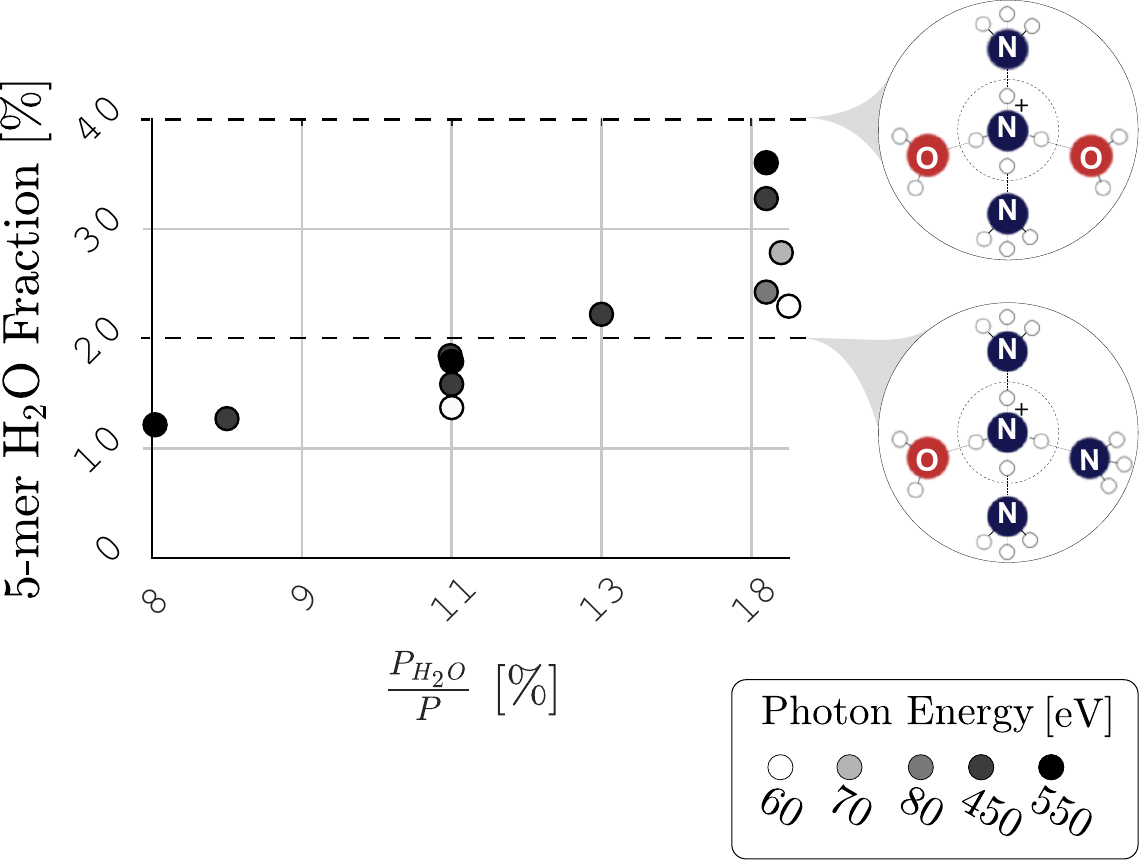}}
\caption{Average water fraction of pentamer units \added[id=MG]{($\ce{(NH_4)^+(NH_3)_{x} (H_2O)_{4-x}}$ where x$\leq$ 4)} as function of the relative water pressure, at photon energies 60, 70, 80, 450 and 550 eV. The right insets show an illustration of the corresponding core-ligand structures.}
\label{fig:fragment}
\end{figure}

As noted earlier, the water fraction (Fig. \ref{fig:dication} a) also exhibits a dependence on the stagnation mixture, increasing from nearly water-free clusters at an 8\% water stagnation pressure, to up to 80\% water in a cluster at only 20\% water partial pressure. This water `enrichment' has been observed in mixed van der Waals clusters through adiabatic co-expansion\cite{Konotop2015, Danilchenko_2011}, and is expected to be due to the difference in the condensation probability between gases. Although the critical variables are still being debated\cite{Konotop2015}, the difference in dimer binding energy (0.163 eV  for water \cite{Barnett1997} vs 0.134 eV for ammonia\cite{Beu2001}) can imply that water molecules condense more readily. In this case, water will mostly be situated in the center of the neutral cluster, in line with previous studies on water-vapor interfaces \cite{Paul_2005, Carignano_2008}.

This water enrichment mechanism was confirmed by the theoretical results. Different cluster sizes with different water-ammonia ratios have been taken into consideration. In all cases, ammonia molecules remain bonded in the outer layer while water molecules clusterize inside, in a sphere-like structure, as the stability of the neutral cluster increases with the water fraction, see Fig. \ref{fig:str10}, \ref{fig:str24} and \ref{fig:str50}. This can also be understood in terms of the strength of the different hydrogen bonds (HB) in the system. Piekarski et. al. \cite{Piekarski_2017}, using wave function analysis methods, studied the strength of the HBs involved in the stabilization of $\beta$-alanine clusters and found that higher electron densities appeared in those HBs where oxygen acts as the donor atom. Therefore, in agreement with a molecular orbital description, the more electronegative the hydrogen donor atom, the greater the stabilization effect. Hydrogen bonds that participate the most in the stabilization of the cluster follow the subsequent order; O$-$H$\cdot \cdot \cdot$N $>$ O$-$H$\cdot \cdot \cdot$O $>$ N$-$H$\cdot \cdot \cdot$N $>>$ N$-$H$\cdot \cdot \cdot$O. This explains our result, where in the formation of the cluster, water remains in the core and ammonium molecules bond to its surface.

The dication composition and critical size do not strongly depend on the photon energy, over the range of this study (see Fig. \ref{fig:dication} and \ref{fig:fragment}). However, a slight overall increase of the water fraction is observed at high photon energies.
It can be argued that this is an effect of increasing internal energy deposited by the double ionization. More energy transferred from the initially electronically excited state, or Coulomb potential energy, to the cluster vibrational, rotational and vibrational motion, would lead to an overall increase of cluster `temperature'. 
This would enhance the evaporation rate at the surface, where the ammonia species are expected to be located. The higher evaporation rate of ammonia with respect to water indeed corresponds to a lower evaporation energy from pure ammonia clusters with respect to the water case \cite{Nakai2000201, MAGNERA1991363}, in line with the water concentration increase at higher photon energies.

In summary, in mixed water-ammonia doubly charged clusters the distance between the two freely moving charges in the dication will be maximized by moving close to the surface\cite{Shukla1984, Harris_2017}. Based upon the condensation conditions, the ammonia molecules are most likely located at the surface. In some cases, the ionization of two nearby ammonia molecules leads to charge separation and the formation of unprotonated species (section \ref{sec:few}). In other cases, the preference of the charges to be located around ammonium and ammonia could result in an increased local concentration of water in between charge centers (section \ref{sec:penta}). Water thus becomes an important stabilizer of the cluster due to water's higher binding energy and higher bulk permittivity ($\tfrac{\epsilon_{\ce{H_2O}}}{\epsilon_{\ce{NH_3}}} \approx 3.1$ \cite{Echt_1988}), which balance the Coulomb repulsion between the two repelling ammonium ions. The stability of the dication clusters will thus strongly depend upon the water composition in the cluster; these observations have been confirmed by theoretical simulations \added[id=SDT]{(see Fig. \ref{fig:energies})}.

\subsection{Nucleation shell}
\label{sec:penta}

We now discuss fragments that have been detected as a (coincident) correlated and completely protonated pair. These fragments are observed for all combinations of fragment size, from symmetric (equal) to asymmetrically sized pairs.
Ammonium ions in hydrated ammonia clusters have often been described in the literature as having the tendency to form shell structures of five molecular units \cite{Hogg_1965,Payzant_1973}. Under all experimental conditions covered in this work, the fragments are observed to host at least one ammonium as charge carrier \cite{Hvelplund_2010} surrounded preferentially by ammonia molecules. \added[id=MG]{We focus our study on fragments with a `filled' first shell around the ammonium (pentamer unit: $\ce{(NH_4)^+(NH_3)_{x} (H_2O)_{4-x}}$ where x$\leq$ 4), as a function of the water pressure ratio and photon energy.}

In Fig. \ref{fig:fragment} the average water fraction of the protonated pentamer unit is shown (left) together with an illustration of a corresponding core-ligand structure (right). The water content appears to only influence the composition of the ligands around \ce{NH}$_4^+$, confirming the competition between ammonia and water to occupy ligand sites \cite{Hvelplund_2010}. At higher photon energies, the water fraction of the pentamer unit increases slightly (see Fig. \ref{fig:fragment}).
Upon core ionization, more energy can be deposited in the cluster, for example by accessing higher electronic states. Ionization at different sites (O1s at 550 eV or N1s at 450 eV) results in a small difference in size or composition, indicating that the internal energy is essentially statistically redistributed into all nuclear degrees of freedom (e.g. during proton transfer \cite{Cheng_1996}), leading to an increase of the relative motions in the cluster and a decrease in the fragment size \cite{Last_2005}.

As mentioned, ammonium prefers ammonia over water ligands to form a shell-like structure,\cite{Choo_1983, Payzant_1973} but as the relative molecular motions intensify, the `propensity' of molecules to occupy the ligand site might randomize to a larger degree, increasing the probability of water molecules from the water-rich cluster environment occupying the ligand site.

Even though the parent cluster composition is not directly measured, it is interesting to relate the stable dication composition to the water fraction in the fragments measured under the same conditions. For instance, in Fig. \ref{fig:comp}, we observe a stable dication water fraction of more than 70 percent, while the fragments contain less than 25 percent water. The large discrepancy in composition suggests a dynamic `selection' of ligand molecules in the ammonium-centered fragment upon dissociation, even at high water fraction in the mixed clusters (or relative water pressure) \cite{Hogg_1965}. An alternative explanation could be the production of a wide distribution of cluster compositions upon adiabatic expansion, but this is unlikely according to our theoretical calculations (see Fig. \ref{fig:str10}, \ref{fig:str24} and \ref{fig:str50}).

\subsection{Molecular interactions}
\label{sec:few}

\begin{figure}[t]
\centerline{\includegraphics[width=0.45\textwidth]{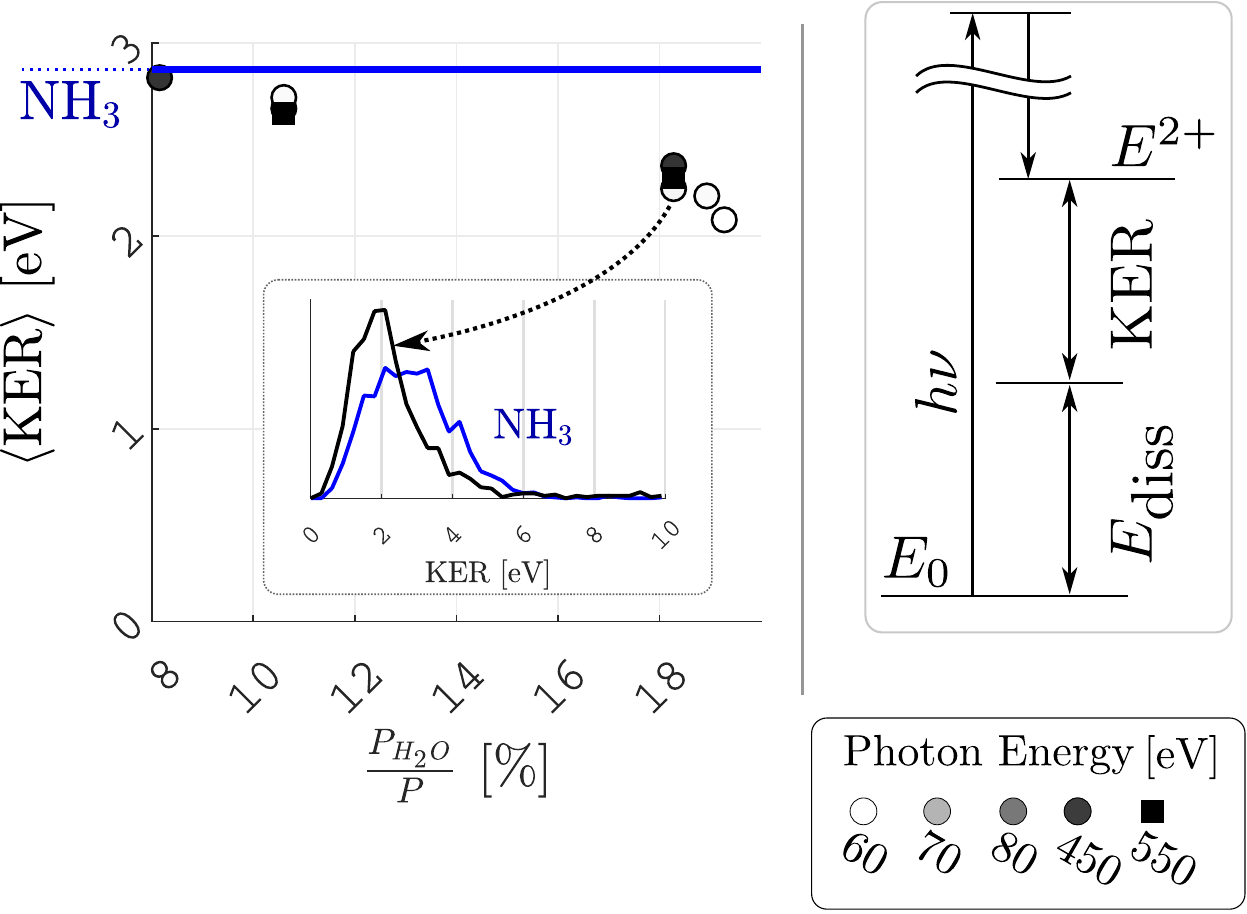}}
\caption{The average Kinetic Energy Release (KER) of $\rm NH_3^+$ and $\rm (NH_3)H^+$ (or $\rm H_2O^+$) at different water pressure ratios and photon energies of 60, 70, 80, 450 and 550 eV, with fragments from pure $\rm NH_3$ clusters as a reference (blue) line. Inset left: two examples of KER distributions at 18\% $\rm H_2O$ mixing and 60, 70, 80, 550 eV photon energy (black line) and pure $\rm NH_3$ (blue). Inset right: schematic energy diagram. Grayscale filling of the datapoints indicates photon energy. The total fragment momentum is restricted ($<$ 60 a.u.) to ensure completeness of the analysis \cite{Oostenrijk_2018}.}
\label{fig:unprot}
\end{figure}

\begin{figure*}
\centering
 \includegraphics[width=1\textwidth]{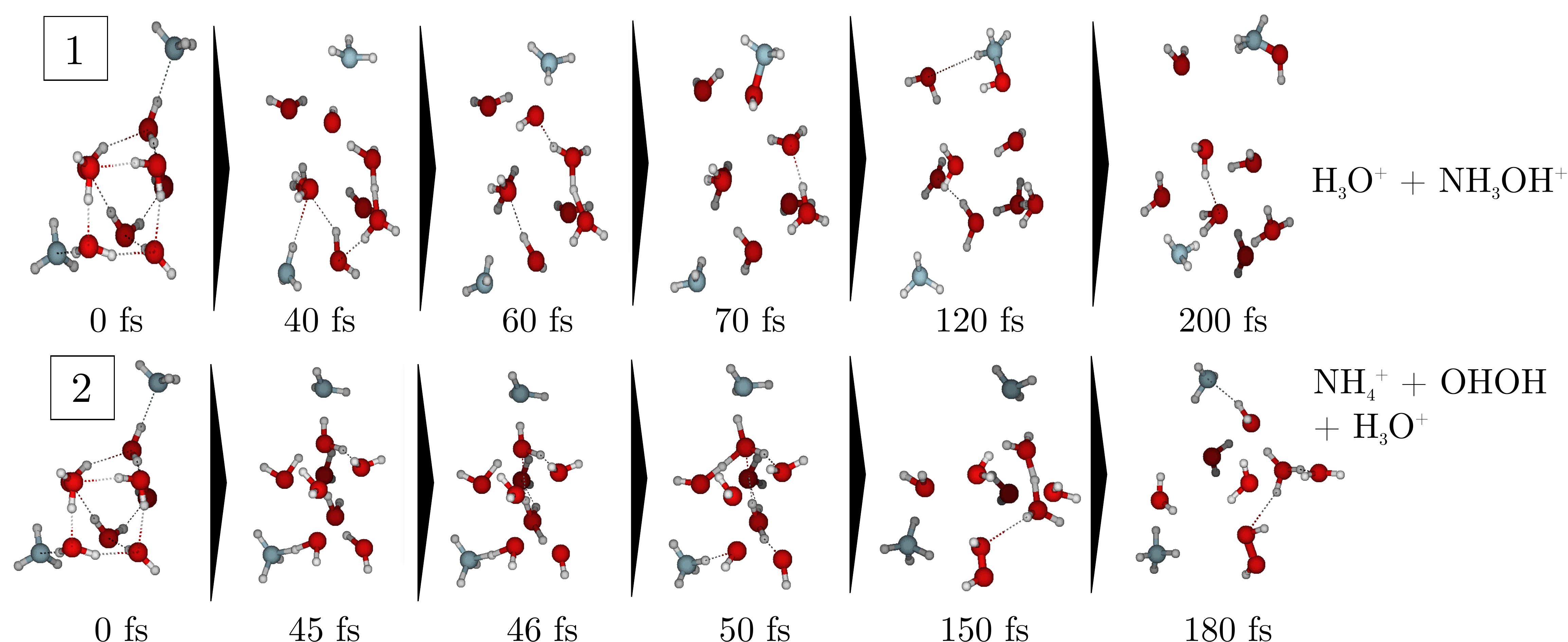}
\caption{Snapshots of the molecular dynamics simulations of the two more representative reactive exit channels, referred as 1 and 2, leading to the fission of the most stable conformer $\rm (NH_3)_2(H_2O)_8$ cluster as a function of time.}
\label{fig:snap1}
\end{figure*}
\begin{figure*}
\centering
 \includegraphics[width=1\textwidth]{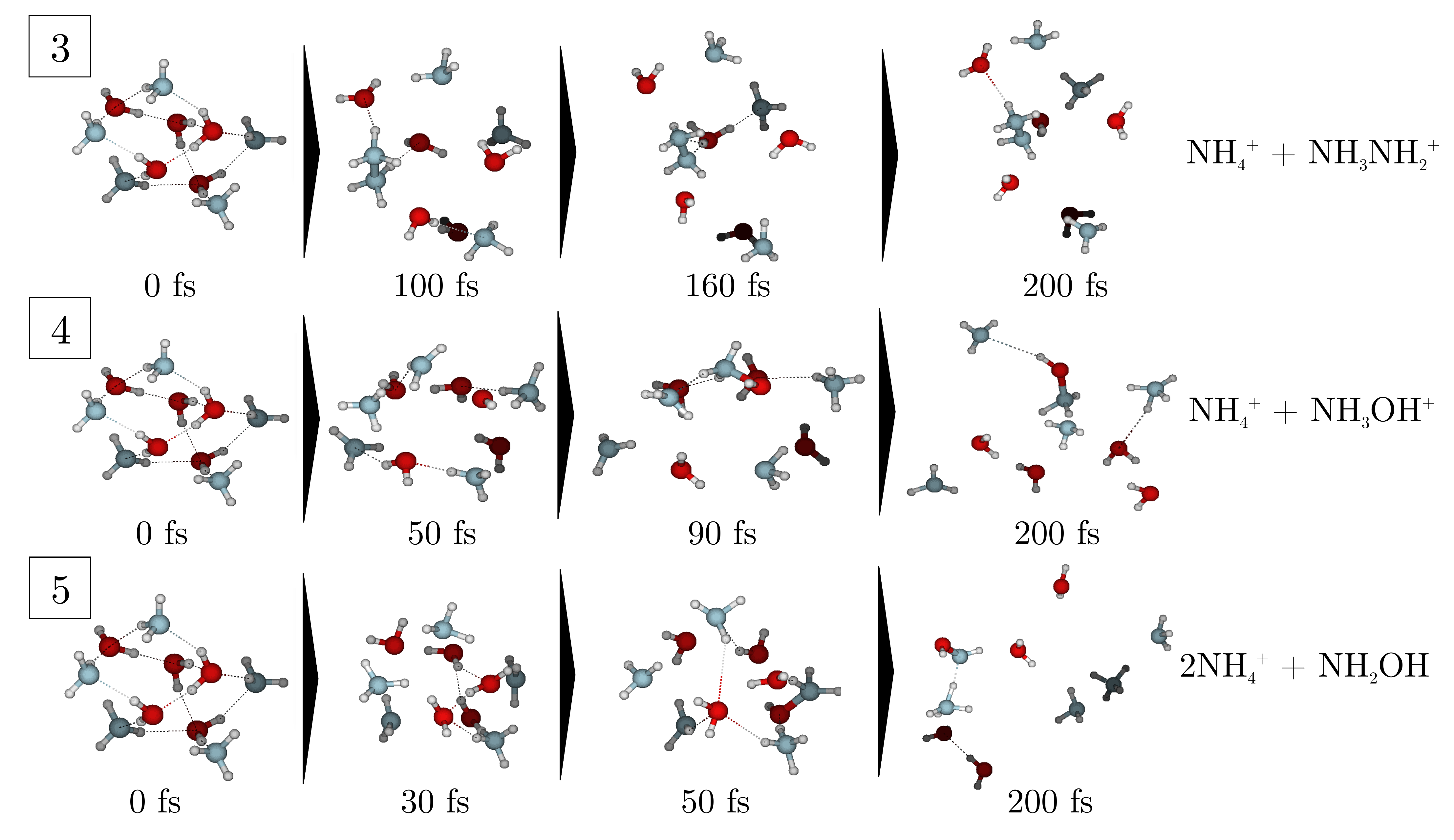}
\caption{Snapshots of the molecular dynamics simulations of the three more representative reactive exit channels, referred as 3, 4 and 5, leading to the fission of the most stable conformer $\rm (NH_3)_5(H_2O)_5$ cluster as a function of time.}
\label{fig:snap2}
\end{figure*}
\begin{figure*}
\centering
 \includegraphics[width=0.9\textwidth]{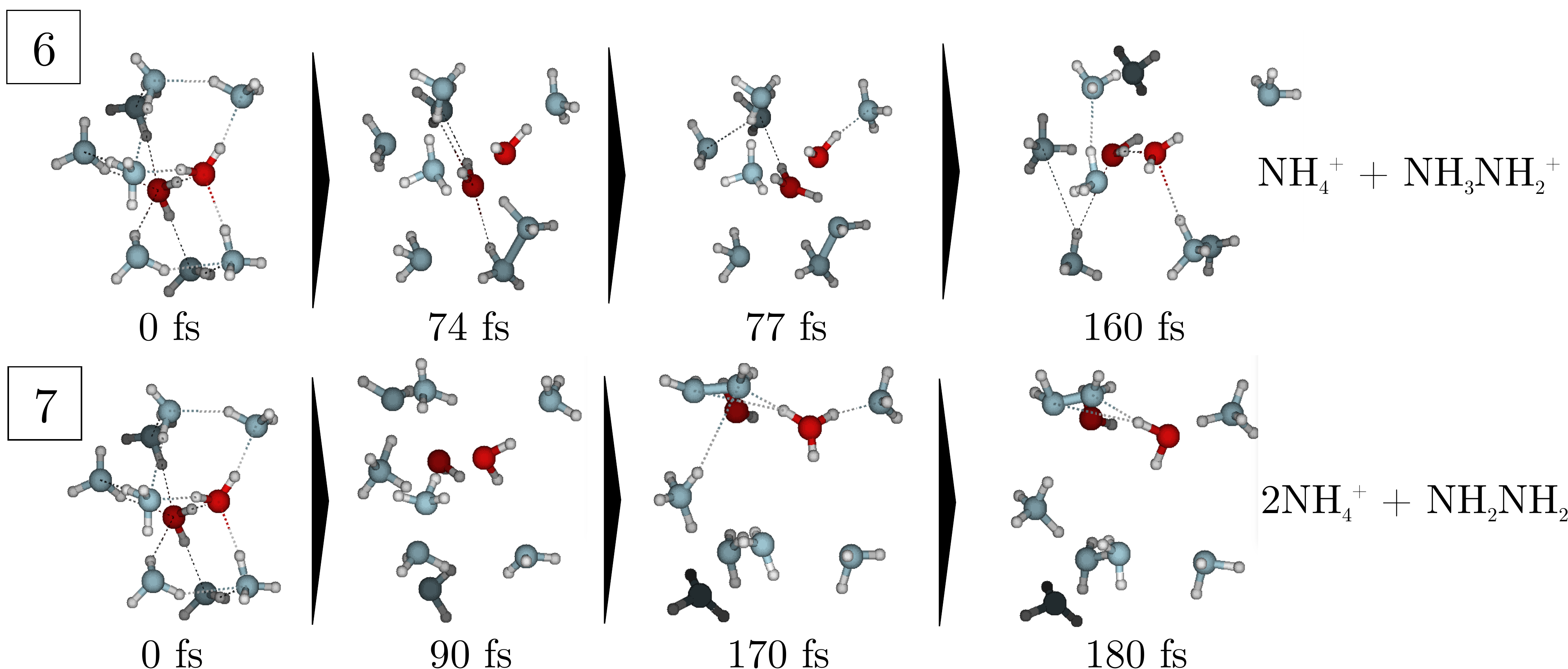}
\caption{Snapshots of the molecular dynamics simulations of the two more representative reactive exit channels, referred as 6 and 7, leading to the fission of the most stable conformer $\rm (NH_3)_8(H_2O)_2$ cluster as a function of time.}
\label{fig:snap3}
\end{figure*}

Inner shell ionization is primarily localized, and the charge and energy diffuses with time around the ionization site, increasing the number of interacting molecules. At short time-scales, the interaction is thus spatially confined near the site of ionization, and can be identified by the production of reactive species, such as unprotonated fragments, i.e. before a proton has transferred to surrounding molecules in the cluster \cite{Oostenrijk_2018}. In the present experiment, we observe a channel consisting of such unprotonated fragment(s); NH$_3^+$ + (NH$_3$)H$^+$ (or NH$_3^+$ + H$_2$O$^+$). In Fig. \ref{fig:unprot} we show the average Kinetic Energy Release (KER) of this fragmentation channel. 
The average KER decreases by 0.7$\pm$0.2 eV as the water content in the environment increases. The mean KER is much lower than the typical KER upon breakup of the water or ammonia dication dimers ($\sim$ 4 eV \cite{Jahnke2010,Kryzhevoi_2011}), and the decrease in kinetic energy (compared to a dimer dication breakup) suggests thus that the dissociation mechanism involves more than two molecules, where the undetected species carries away energy from both detected fragments. 
The schematic energy diagram in Fig. \ref{fig:unprot} highlights the relevant energy levels. Upon ionization, doubly charged states are prepared ($E^{2+}$) which undergo dissociation, releasing kinetic energy to fragments. Depending on the energy of dissociation ($E_{\text{diss}}$), the process will be more or less exothermic. The water-ammonia bond is stronger than the typical ammonia-ammonia bond (about 0.7 eV difference \cite{Jeffrey_H_bond}), which is the same approximate difference in the observed kinetic energy release of a water-ammonia cluster compared to that of a pure ammonia cluster. The energy difference can thus be due to fragments emitted from an increasingly strong H-bonded lattice/network containing an increasing amount of water molecules. Note that this interpretation is based on the assumption that no other ionization channels are opened by mixing molecules, such as Inter Molecular Coulombic  Decay\cite{Jahnke2010,Kryzhevoi_2011}, with a lower $E^{2+}$ energy level. 

It is worth mentioning that the relative abundance of the discussed dissociation channel shows no dependence on the neutral cluster size distribution, the water pressure fraction or photon energy. The change of stagnation pressure over such a large range is expected to result in a large change in average cluster size \cite{Bobbert_2002}. This will drastically change the production of small neutral clusters, excluding the small (trimer, tetramer, pentamer, etc) parent clusters from being the parent cluster of the fragmentation channel discussed above, and we thus conclude that this breakup channel likely occurs near the surface of the cluster.

\begin{figure*}[ht!]
\centering
 \includegraphics[width=0.9\textwidth]{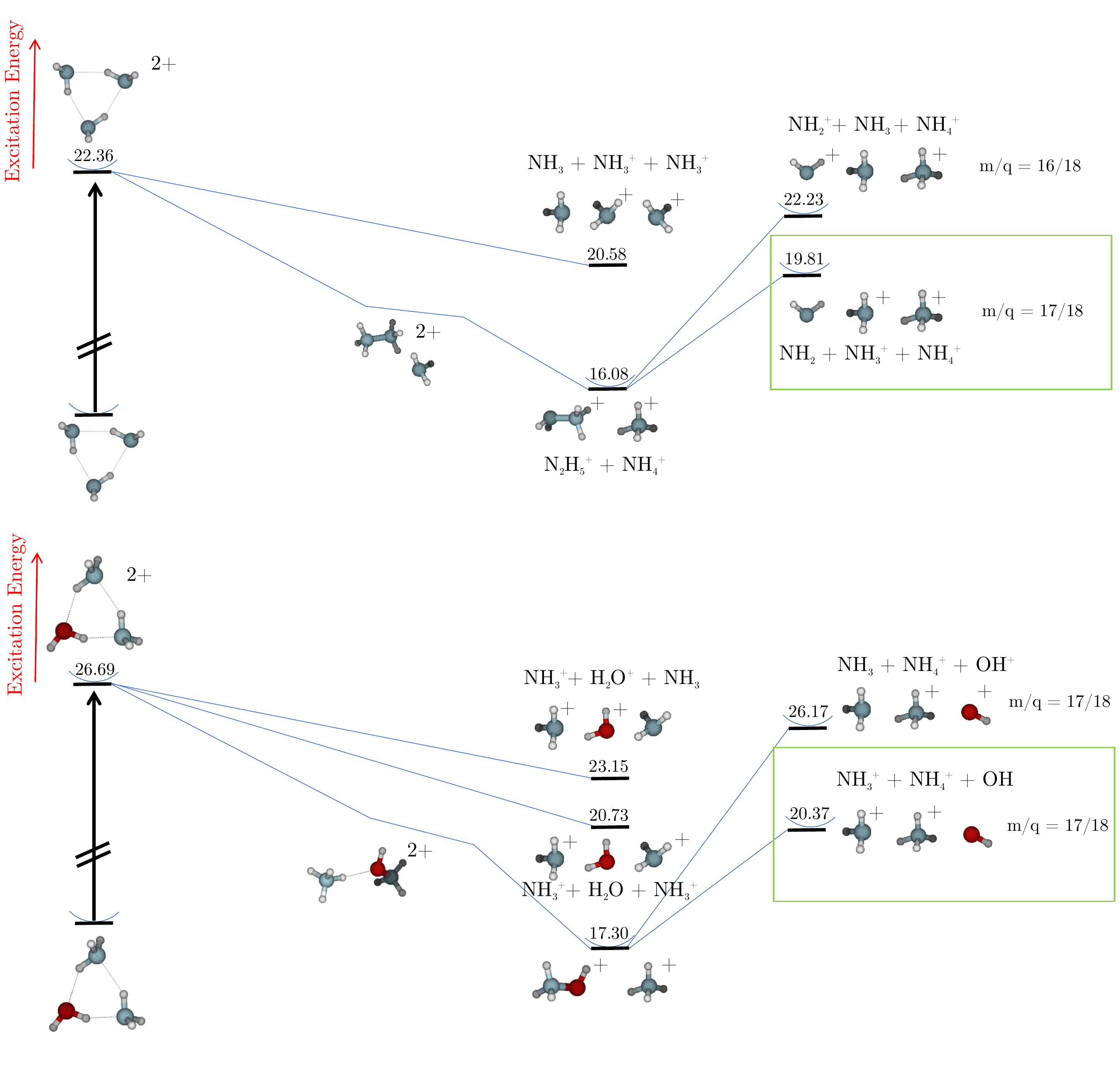}
 \caption{Schematic representation of the potential energy surface exploration for $\rm (NH_3)_3$ and $\rm (NH_3)_2(H_2O)$ after double ionization and posterior dissociation. Relative energies are shown in eV with respect to the most stable neutral dimer. All calculations performed at a DFT level of theory using M06-2X/6-31++(d,p)}
\label{fig:paths}
\end{figure*}

In order to uphold these observations, we take a closer look at the underlying mechanisms taking place at a molecular level during the first femtoseconds after ionization. Hence, ab initio molecular dynamics simulations and, in our case, semi-classical Atom-centered Density Matrix Propagation (ADMP) simulations, allow us to add a theoretical background to experimental results. Once the conformational study and the VIPs were calculated, using them as a starting point, 100 trajectories were performed at 10 and 20 eV of excitation energy at a DFT level of theory for each doubly-charged 10 unit mixed clusters (NH$_3$)$_m$(H$_2$O)$_n$ with ($m,n$) = (2,8), (5,5) and (8,2). The complete result after the statistical study over all the trajectories is presented in the Appendix, while here only the most relevant reactive exit channels for each nano-hydrated cluster will be highlighted. Interestingly, around 50$\%$ of the trajectories performed at the higher excitation energy, 20 eV and around 20$\%$ of the trajectories for 10 eV, are non reactive, therefore dominated by the electrostatic forces and leading to the explosion of the cluster without any hydrogen transfer. Fig. \ref{fig:snap1}, \ref{fig:snap2} and \ref{fig:snap3} show snapshots of some selected trajectories corresponding to the statistically more relevant reactive exit channels for the different 10 unit mixed clusters considered. As can be seen, even in the case of the cluster with higher water fraction the formation of charged NH$_4^+$ is favored over H$_3$O$^+$, which bears out the experimental results. In these simulations, charge is localized preferably on the amino compounds while, on the other hand, water molecules act as 'proton jumping intermediates' as in the Grotthuss mechanism\cite{Grothuss_1805}. 
At the end of the 300 fs simulation, the system evolves  along different pathways leading to the detected exit channels. In the case of the non reactive path, the Coulomb explosion, each charge is localized  on an ammonia molecule.  Despite experimental limitation in detecting particles at rest with the same mass, experiments show no record of fragment pair with kinetic energies with masses 17 and 17 in double coincidence spectra. Therefore, it is possible that the excitation energy assimilated by the cluster upon ionization is closer to 10 eV than to 20 eV and that the system is very likely evolving through a reactive channel after the first 300 fs and once the Coulombic explosion has waned its effect. In the present experiments, the signal of fragment pair with masses 17 and 18 was highlighted and assigned to NH$_4^+$ (or H$_2$O$^+$)/NH$_3^+$. Energetically, the formation NH$_4^+$ will be strongly favored over H$_2$O$^+$. In fact, NH$_4^+$/NH$_3^+$ is the most energetically favorable evolution for the reactive exit channels 3, 4 and 6, with these being the statistically most populated pathways for those clusters with higher ammonia fractions. Fig. \ref{fig:paths} shows the evolution after formation of a positive doubly-charged (NH$_3$)$_3^{2+}$ cluster. As can be seen, instead of the reorganization of the charge on isolated ammonia molecules, the lower reaction path involves the barrier-less formation of an ammonia dimer and a hydrogen transfer leading to the posterior break-up of the bond and the formation of neutral NH$_2$ and singly-charged NH$_4^+$ and NH$_3^+$. Also, it shows the most probable reaction path for those clusters with higher water fractions. In this case a cluster consisting of two ammonia molecules and one water (NH$_3$)$_2$(H$_2$O) have been taken into consideration. Here, after ionization, the doubly-charged cluster evolves through the formation of a N-O bond and a hydrogen transfer, producing NH$_4^+$ and \ce{NH_3OH}$^{+}$. From here, the most probable evolution involves the break-up of the N-O bond leading to a final state with neutral OH and singly-charged NH$_4^+$ and NH$_3^+$.

\section{Conclusions}
To bridge the gap between the few molecule response and the nanoscale nucleation process of aerosols, we studied fragmentation of water-ammonia clusters produced via adiabatic co-expansion using soft X-ray radiation and quantum chemistry simulations. By analyzing the critical stable size of doubly charged clusters, we found a water `enriched' composition, compared to the stagnation pressure ratio, which can be explained by the higher condensation rate of water upon adiabatic expansion.

Water molecules are recognized as an effective stabilizer of the doubly charged cluster, since it enables to host two charges at a nearly pure water nano-droplet size with just 30\% water content in the cluster. Similarly, on the molecular scale, the interaction of ammonia molecules with an increasing number of surrounding water molecules leads to a decrease in the kinetic energy released in an exothermic reaction, suggesting that the energy required to dissociate from the H-bonded lattice/network increases. The simulations confirm the stabilizing influence of water molecules when they act as 'core molecules' in nucleation, with the ammonia remaining in the outer shells.

Finally, in agreement with previous studies and regardless of the water concentration and initial charge states, a core-shell structure around ammonium is always determining for the composition of small fragments. We conclude that a dynamic selection of ligand molecules occurs, and leads to ammonia-abundant protonated small fragments. In this core-shell structure, the aqueous ammonia seems to act as an effective proton `collector', forming a central ammonium, while water molecules act as 'proton tunneling' intermediates in a Grotthus-like mechanism. \textit{Ab initio} molecular dynamics simulations provide insights on the dynamics of such processes. The dominance of ammonium as a charge carrier, already at a high water content in the cluster, indicates that it has the ability to act as a reaction center for chemical activity with known consequences on the nucleation rates in the atmosphere \cite{Ball1999,Torpo2007,Almeida2013}.

\section*{Conflicts of interest}
There are no conflicts to declare.

\section*{Acknowledgements}
M.G., N.W, and B.O would like to thank  Maxim Tchaplyguine for his support during the experiment. B.O acknowledges the grant associated to this project by the Crafoord foundation (reference 20180946). We acknowledge the generous allocation of computer time at the Centro de Computaci\'on Cient\'{\i}fica at the Universidad Auton\'oma de Madrid (CCC-UAM).
This work was partially supported by the project CTQ2016-76061-P of the Spanish Ministerio de Econom\'{\i}a y Competitividad (MINECO).
D.B. acknowledges the FPI grant associated to this project (grant reference BES-2017-080127).
Financial support from the MINECO through the ``Mar\'{\i}a de Maeztu'' Program for Units of Excellence in R\&D (MDM-2014-0377) is also acknowledged.
The authors gratefully acknowledge J\'er\^{o}me Cuny and Mathias Rapacioli for technical help with the DFTB simulations using the deMonNano code.



\balance


\bibliography{rsc} 
\bibliographystyle{rsc} 

\newpage

\section*{Appendix}


\begin{table}[h!]
\small
  \label{tab:VIP}
  \begin{tabular}{rcc||rcc}
\hline
(NH$_3$)$_m$ & 1$^{\rm st}$+2$^{\rm nd}$ VIP & DE & (H$_2$O)$_n$ & 1$^{\rm st}$+2$^{\rm nd}$ VIP & DE \\
\hline
 $m=$1   & 35.91  & -      &$n=$1 & 41.70 & -     \\
     2   & 29.25  & -      & 2   & 34.39 & -     \\
     3   & 26.94  & 24.39  & 3   & 32.16 & 24.51 \\
     4   & 25.06  & 26.12  & 4   & 30.83 & 25.00 \\
     5   & 24.70  & 27.88  & 5   & 29.45 & 25.35 \\
     6   & 24.30  & 29.73  & 6   & 28.76 & 25.72 \\
     7   & 23.87  & 31.49  & 7   & 28.55 & 26.30 \\
     8   & 23.75  & 33.35  & 8   & 28.35 & 26.95 \\
     9   & 22.91  & 35.16  & 9   & 27.99 & 27.33 \\
    10   & 22.56  & 37.00  & 10  & 27.77 & 27.81 \\
\hline
  \end{tabular}
  \caption{Energy data for pure NH$_3$ and H$_2$O clusters computed at the M06-2X/6-31++G(d,p) level of theory.
  Dissociation energies (DE) and Second vertical ionization potentials (VIPs), both in eV.
  In the case of the ammonia clusters, the dissociation channel considered was ${\rm (NH_3)_m \rightarrow NH_4^+ + NH_2 + NH_3^+ + (m-3)NH_3}$, meanwhile, for the water cluster the dissociation channel considered was ${\rm (H_2O)_n \rightarrow H_3O^+ + H_2O^+ + OH + (n-2)H_2O}.$}
\end{table}

\begin{table*}[h!]
\small
  \label{tab:VIP}
  \begin{tabular}{rcc||rcc}
\hline
$m+n=$10 & 1$^{\rm st}$+2$^{\rm nd}$ VIP  & DE & $m+n=$24 & 1$^{\rm st}$+2$^{\rm nd}$ VIP & DE \\
\hline
     m, n =2, 8 & 26.11  & 20.43  & n, m=6, 18 & 23.42 & 29.38     \\
     5, 5   & 24.69  & 22.72      & 11, 12   & 22.36 & 28.22     \\
     8, 2   & 22.88  & 22.15  & 18, 6    & 21.16 & 27.17 \\
\hline
  \end{tabular}
  \caption{Energy data for 10-units and 24-units mixed ${\rm (NH_3)_m(H_2O)_n}$ clusters computed at the M06-2X/6-31++G(d,p) level of theory.
   Dissociation energies (DE) and Second vertical ionization potentials (VIPs), both in eV.
   The dissociation channel taken into consideration is:
${\rm (H_2O)_y(NH_3)_x \rightarrow y H_2O + NH_4^+ + NH_2 + NH_3^+ + (x-3)NH_3}$
except for ${\rm (H_2O)_8(NH_3)_2 \rightarrow 7 H_2O + NH_4^+ + NH_3OH^+}$.}
\end{table*}


\begin{figure*}[h!]
\centering
 \includegraphics[width=0.6\textwidth]{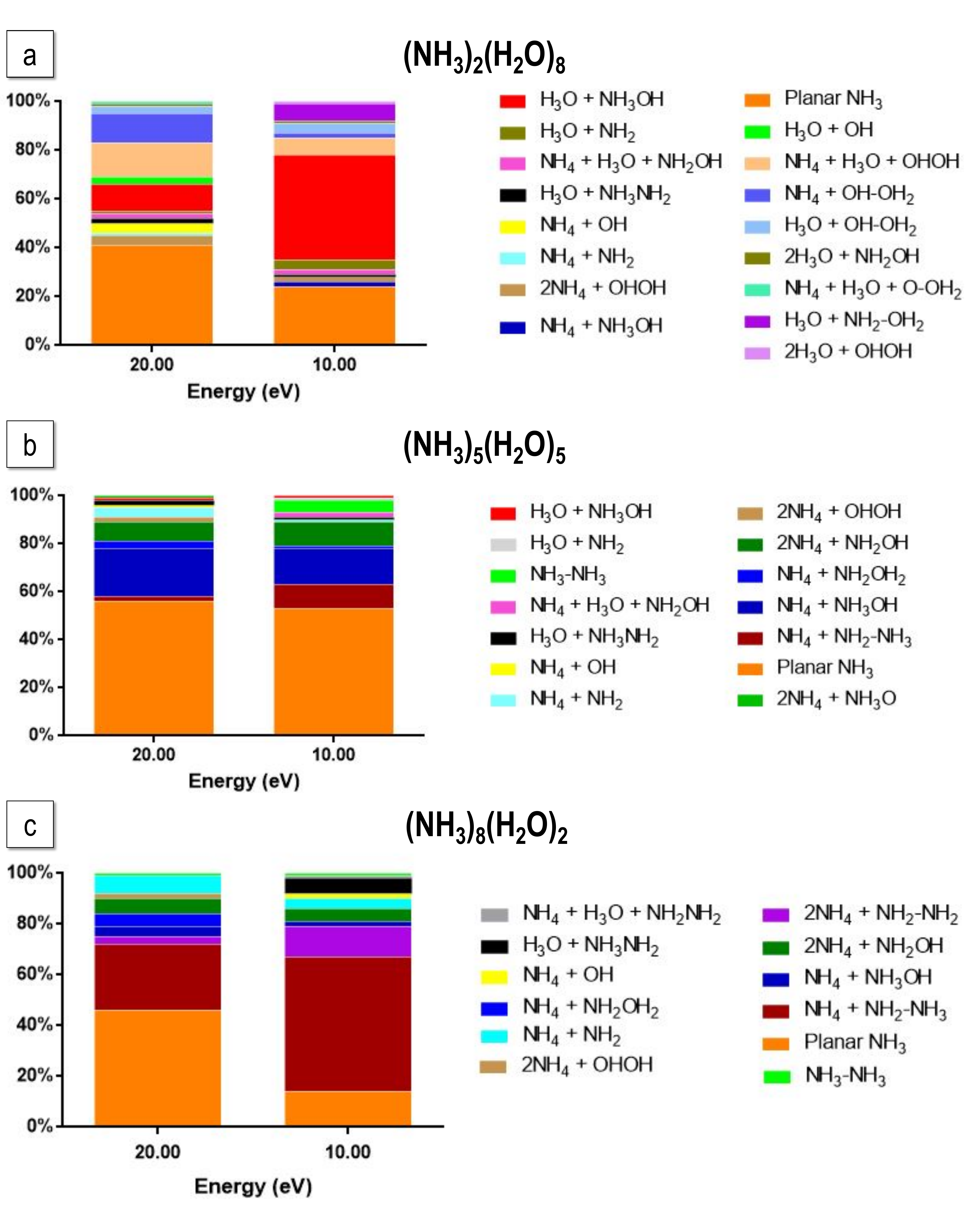}
 \caption{{\it Ab initio} molecular dynamics statistics for the most stable conformer of (a): (NH$_3$)$_2$(H$_2$O)$_8$, (b): (NH$_3$)$_5$(H$_2$O)$_5$ and (c): (NH$_3$)$_8$(H$_2$O)$_2$.}
\label{fig:app}
\end{figure*}

\end{document}